\documentclass[floatfix,onecolumn,showpacs]{revtex4}

\usepackage{graphics,graphicx}
\usepackage{epsfig}
\usepackage{amsmath}
\usepackage{amstext}
\usepackage{amssymb}
\usepackage{dcolumn}

\newcommand{\beq}{\begin{eqnarray}}
\newcommand{\eeq}{\end{eqnarray}}

\begin{document}

\title{Towards a large deviation theory for statistical-mechanical complex systems}

\author{Guiomar Ruiz$^{1,2}$  and Constantino Tsallis$^{1,3}$}
\affiliation{$^1$Centro Brasileiro de Pesquisas Fisicas and \\National Institute of Science and Technology for Complex Systems,
Rua Xavier Sigaud 150, 22290-180 Rio de Janeiro-RJ, Brazil\\
$^2$ Departamento de Matem\'{a}tica Aplicada y Estad\'{\i}stica, Universidad Polit\'{e}cnica de Madrid, Pza. Cardenal Cisneros s/n, 28040 Madrid, Spain\\
$^3$Santa Fe Institute, 1399 Hyde Park Road, Santa Fe, NM 87501, USA}

\begin{abstract}
The theory of large deviations constitutes a mathematical cornerstone in the foundations of Boltzmann-Gibbs statistical mechanics, based on the additive entropy $S_{BG}=- k_B\sum_{i=1}^W p_i \ln p_i$. Its optimization under appropriate constraints yields the celebrated BG weight $e^{-\beta E_i}$. An elementary large-deviation connection is provided by $N$ independent binary variables, which, in the $N\to\infty$ limit yields a Gaussian distribution. The probability of having $n \ne N/2$ out of $N$ throws is governed by the exponential decay $e^{-N r}$, where the rate function $r$ is directly related to the relative BG entropy.  To deal with a wide class of complex systems, nonextensive statistical mechanics has been proposed, based on  the nonadditive entropy $S_q=k_B\frac{1- \sum_{i=1}^W p_i^q}{q-1}$ ($q \in {\cal R}; \,S_1=S_{BG}$). Its optimization yields the generalized weight $e_q^{-\beta_q E_i}$ ($e_q^z \equiv [1+(1-q)z]^{1/(1-q)};\,e_1^z=e^z)$. We numerically study large deviations for a strongly correlated model which depends on the indices $Q \in [1,2)$ and $\gamma \in (0,1)$. This model provides, in the $N\to\infty$ limit ($\forall \gamma$), $Q$-Gaussian distributions, ubiquitously observed in nature ($Q\to 1$ recovers the independent binary model).  We show that its corresponding large deviations are governed by $e_q^{-N r_q}$ ($\propto 1/N^{1/(q-1)}$ if $q>1$) where $q= \frac{Q-1}{\gamma (3-Q)}+1 \ge 1$. This $q$-generalized illustration opens wide the door towards a desirable large-deviation foundation of nonextensive statistical mechanics.
\end{abstract}
\pacs{02.50.-r,05.20.-y,05.40.-a,65.40.gd}

 \maketitle

In his 1902 historical book {\it Elementary Principles in Statistical Mechanics} \cite{Gibbs1902}, Gibbs emphasizes that systems involving long-range interactions are intractable within the Boltzmann-Gibbs (BG) theory, due to the divergence of the partition function. Amazingly enough, this crucial remark is often overlooked in most textbooks. However, this is clearly why no standard temperature-dependent thermostatistical quantities (e.g., specific heat) are computable for the free hydrogen atom, for instance. Indeed, an infinite number of excited energy levels accumulate at the ionization value, which makes the canonical partition function to trivially diverge at any finite temperature.

To solve this and related complexities it has been proposed in 1988 \cite{Tsallis1988,GellMannTsallis2004,TsallisGellMannSato2005,Tsallis2009} a generalization of the BG theory, currently referred to as nonextensive statistical mechanics. It is based on the nonadditive entropy
\begin{equation}
S_q=k_B\frac{1-\sum_{i=1}^Wp_i^q}{q-1} \;\;\;\Bigl(q \in {\cal R}; \, \sum_{i=1}^W p_i=1 \Bigr),
\label{sq}
\end{equation}
which recovers $S_{BG}= - k_B\sum_{i=1}^W p_i \ln p_i$ for $q\to 1$. If $A$ and $B$ are two probabilistically independent systems (i.e., $p_{ij}^{A+B}=p_i^Ap_j^B$, $\forall (i,j)$), definition (\ref{sq}) implies the nonadditive relation $\frac{S_q(A+B)}{k_B}=  \frac{S_q(A+B)}{k_B}+ \frac{S_q(A+B)}{k_B}+(1-q)\frac{S_q(A+B)}{k_B}\frac{S_q(A+B)}{k_B}$. Moreover, if probabilities are all equal, we straightforwardly obtain $S_q=k_B \ln_q W$, with $\ln_q z \equiv \frac{z^{1-q}-1}{1-q}$  ($\ln_1 z=\ln z$). If we extremize (\ref{sq}) with a constraint on its width (in addition to normalization of the probabilities $\{p_i\}$), we obtain
\begin{equation}
p_i=\frac{e_q^{-\beta_q\,E_i}}{\sum_{j=1}^W e_q^{-\beta_q\,E_j}} \,,
\label{pq}
\end{equation}
$e_q^z$ being the inverse function of the $q$-logarithmic function, i.e., $e_q^z \equiv [1+(1-q)z]^{1/(1-q)}$ ($e_1^z=e^z$); $\{E_i\}$ are the energy levels; $\beta_q$ is an effective inverse temperature.

Complexity frequently emerges in natural, artificial and social systems. It may be caused by various geometrical-dynamical ingredients, which include non-ergodicity, long-term memory, multifractality, and other spatial-temporal long-range correlations between the elements of the system. During the last two decades, many such phenomena have been successfully approached in the frame of nonextensive statistical mechanics. Predictions, verifications and various applications have been performed in high-energy physics \cite{CMS1,ALICE1,ALICE4,ATLAS,PHENIX1,ShaoYiTangChenLiXu2010}, spin-glasses \cite{PickupCywinskiPappasFaragoFouquet2009}, cold atoms in optical lattices \cite{DouglasBergaminiRenzoni2006}, trapped ions \cite{DeVoe2009},
anomalous diffusion \cite{AndradeSilvaMoreiraNobreCurado2010}, dusty plasmas \cite{LiuGoree2008},  solar physics \cite{BurlagaVinasNessAcuna2006,BurlagaNess2009,BurlagaNessAcuna2009,
ChoLazarian2009,EsquivelLazarian2010}, relativistic and nonrelativistic nonlinear quantum mechanics \cite{NobreMonteiroTsallis2011}, among many others.

It is known since several decades that the mathematical foundation of BG statistical mechanics crucially lies on the theory of large deviations (see \cite{Ellis1985,Touchette2009} and references therein). To attain the same status for nonextensive statistical mechanics, it is necessary to $q$-generalize the large deviation theory itself. The purpose of the present effort precisely is to make a first step towards that goal through the study of a simple model.

\begin{figure}
\begin{center}
\includegraphics[width=11cm]{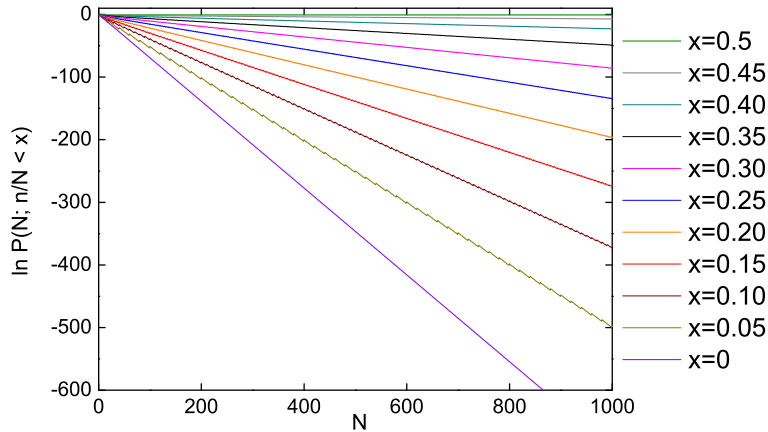}
\end{center}\vspace{-0.5cm}
\caption{Tossing $N$ independent coins: the large-deviation probability $P(N; \,n/N<x)$ decays exponentially with $N$. The slopes are given by the rate function $r_1(x)$, which is shown in Fig. \ref{fig2}
\label{fig1}}
\end{figure}

Let us start with the standard example which consists in tossing $N$ times a (fair) coin. The  probability of obtaining $n$  ($n=0,1,\dots N$) heads  is given by
\begin{equation}
p_{N,n}=\left( \begin{array}{c}N \\  n \end{array}\right)\frac{1}{2^N} \,,
\end{equation}
and the probability of having a ratio $n/N$ smaller than $x$ with $0 \le x \le 1/2$ (the case $1/2 \le x \le 1$ is totally symmetric) is given by
\begin{equation}
P(N; \,n/N<x) =\sum_{\{n | \frac{n}{N}<x\}} p_{N,n}
\label{PN}
\end{equation}
It is straightforward to obtain that, in the $N\to\infty$ limit,
\begin{equation}
P(N; \,n/N<x) \simeq e^{- N\,r_1(x)} \;\;\;(0 \le x \le 1/2)\,,
\label{PN1}
\end{equation}
where the subindex 1 in the {\it rate function} $r_1(x)$ will soon become clear. This exponential decay with $N$, deeply related with the exponential decay with energy of the BG weight (namely, the $q=1$ particular case of Eq. (\ref{pq})), can be verified in Fig.~\ref{fig1}.

\begin{figure}\hspace{-0.8cm}
\begin{center}
\includegraphics[width=9cm]{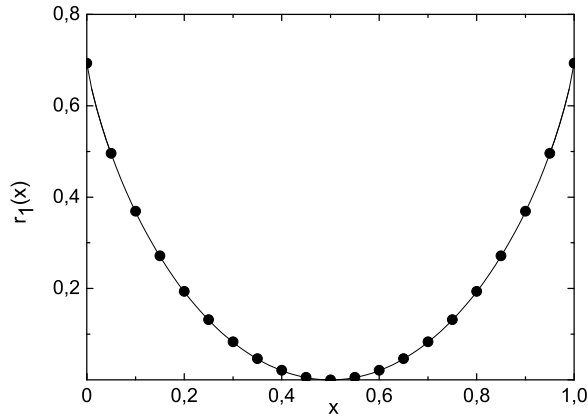}
\end{center}\vspace{-0.5cm}
\caption{The rate function for independent binary variables. The dots have been numerically obtained from Fig. \ref{fig1}. The continuous curve corresponds to Eq. (\ref{r1}).\label{fig2}}
\end{figure}

The rate function $r_1(x)$ also is easy to analytically calculate for the present trivial model.
The {\it relative entropy} or {\it mutual information} for a single random variable with discrete $W$ events with probabilities $\{p_i \}$ ($i=1,\dots, W$) is defined as
\begin{equation}
I_1=-\sum_{i=1}^{W} p_i \ln{\frac{p_i^{(0)}}{p_i}}\label{I_1} \,,
\end{equation}
where $p_i^{(0)}$ is a reference distribution. By choosing $\{p_i^{(0)}\}$ as the uniform distribution, (i.e., $p_i^{(0)}=1/W$) we have
\begin{equation}
I_1=\ln{W}-\sum_{i=1}^{W} p_i \ln{\frac{1}{p_i}}=\ln{W}- \frac{S_1}{k_B}
\end{equation}
For a coin we have $W=2$ ({\it head} or {\it tail}), hence $I_1$ reads
\begin{equation}
I_1=\ln{2}+p_1\ln p_1 +p_2 \ln p_2 \,.
\end{equation}
By identifying $(p_1,p_2) \to (x, 1-x)$ we obtain
\begin{equation}
r_1(x)=\ln{2}+x\ln{x}+(1-x)\ln{(1-x)} \,,
\label{r1}
\end{equation}
as can be verified in Fig. \ref{fig2}.

Before focusing on a model of correlated coins, let us $q$-generalize Eq. (\ref{r1}).
Definition (\ref{I_1}) and the nonadditive entropy $S_q$ naturally lead to the generalization \cite{Tsallis1998}
\begin{equation}
I_q=-\sum_{i=1}^{W} p_i  \ln_q \frac{p_i^{(0)}}{p_i}
=\sum_{i=1}^{W} p_i \frac{[(p_i/p_i^{(0)})^{q-1}-1]}{q-1}\label{I_q}
\end{equation}
By once again choosing as $\{ p_i^{(0)} \}$ the equiprobability distribution (i.e., $p_i^{(0)}=1/W$), we have
\begin{equation}
I_q=W^{q-1}\left[\ln_q{W}-\frac{S_q}{k_B} \right]
\end{equation}
For $W=2$ and $(p_1,p_2) \to (x, 1-x)$, we obtain
\begin{equation}
I_q(x)=\frac{1}{1-q}\left[1-2^{q-1}[x^q+(1-x)^q]\right] \,,
\label{Iqx}
\end{equation}
which recovers expression (\ref{r1}) for $q \to 1$.

Let us now extend the above model to the case where the coins might be strongly correlated. The simple model that we have just reviewed yields, for $N\to\infty$, a Gaussian distribution. This conforms to the Central Limit Theorem, valid for sums of many independent random variables whose variance is finite. This classical theorem has been $q$-generalized \cite{UmarovTsallisSteinberg2008}  for a special class or correlations referred to as $q$-independence ($1$-independence recovers standard independence). The attractors in the probability space are $q$-Gaussians, which precisely extremize the (continuous form of the) entropy $S_q$ when an appropriately generalized $q$-variance is maintained fixed. It is then this class of correlations that we are going to focus on here in order to illustrate how the classical large-deviation theory can be generalized. We adopt the specific class of binary variable models introduced in \cite{RodriguezSchwammleTsallis2008}. These models consist in discretized forms of $Q$-Gaussians ($1 \le Q<2$) \cite{remark}, which exactly converge onto $Q$-Gaussians in the limit $N\to\infty$. The $\{p_{N,n}\}$ in (\ref{PN}) are now given by
\begin{equation}
p_{N,n}= \frac{p_Q(y_{N,n})}{\displaystyle\sum_{n=0}^N p_Q(y_{N,n})} \,,
\end{equation}
where
\begin{equation}
p_Q (z)\propto [1+(Q-1)z^2]^{-1/(Q-1)}
\end{equation}
and
$y_{N,n}$ ($n=0,1,\dots,N$) correspond to $(N+1)$ equally spaced points in the support of the discretized $Q$-Gaussian. More precisely,
$y_{N,n}=\Delta_N\left(\frac{n}{N}-1/2\right) \in[-\Delta_N/2,\Delta_N/2]$, where
\begin{equation}
\Delta_N=\delta (N+1)^\gamma  \;\;\;(\delta>0; \,0< \gamma < 1).
\label{Delta}
\end{equation}

\begin{figure*}
\resizebox{1\columnwidth}{!}{%
  \includegraphics[width=5.3cm,angle=0]{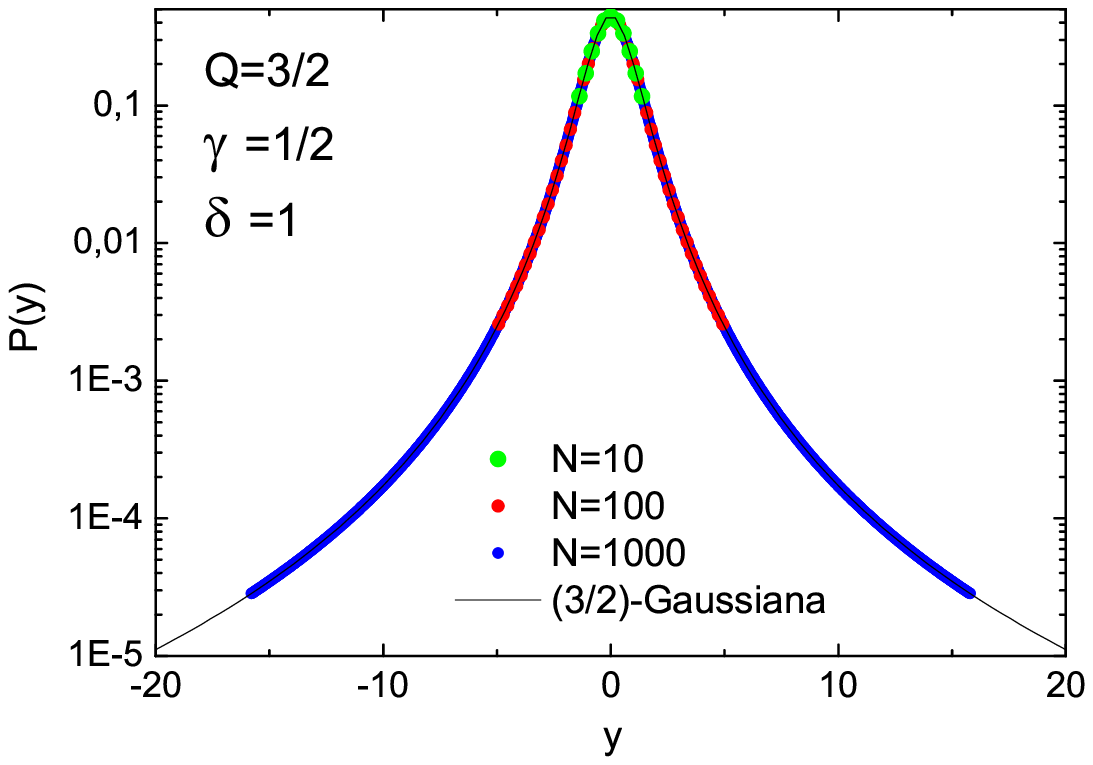}
  \includegraphics[width=6.3cm,angle=0]{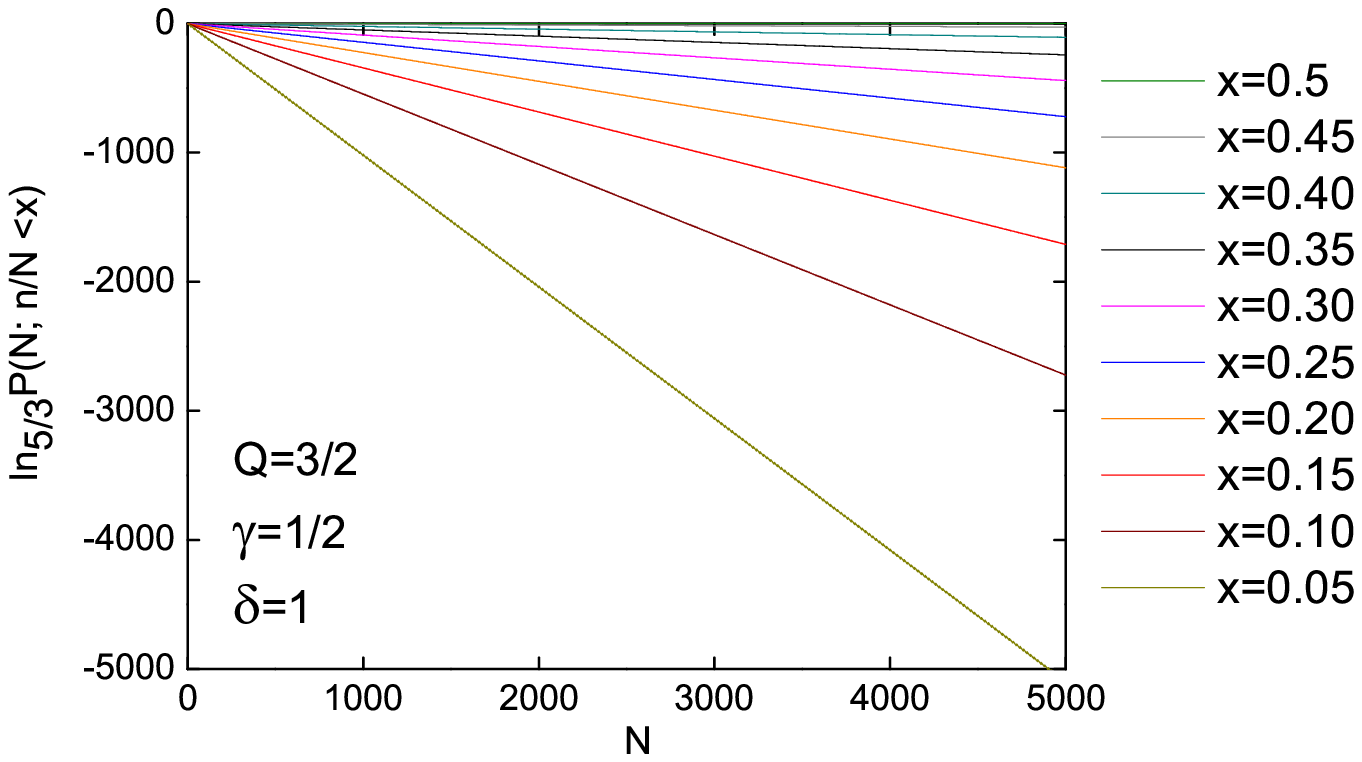}
}
\resizebox{1\columnwidth}{!}{%
  \includegraphics[width=5.3cm,angle=0]{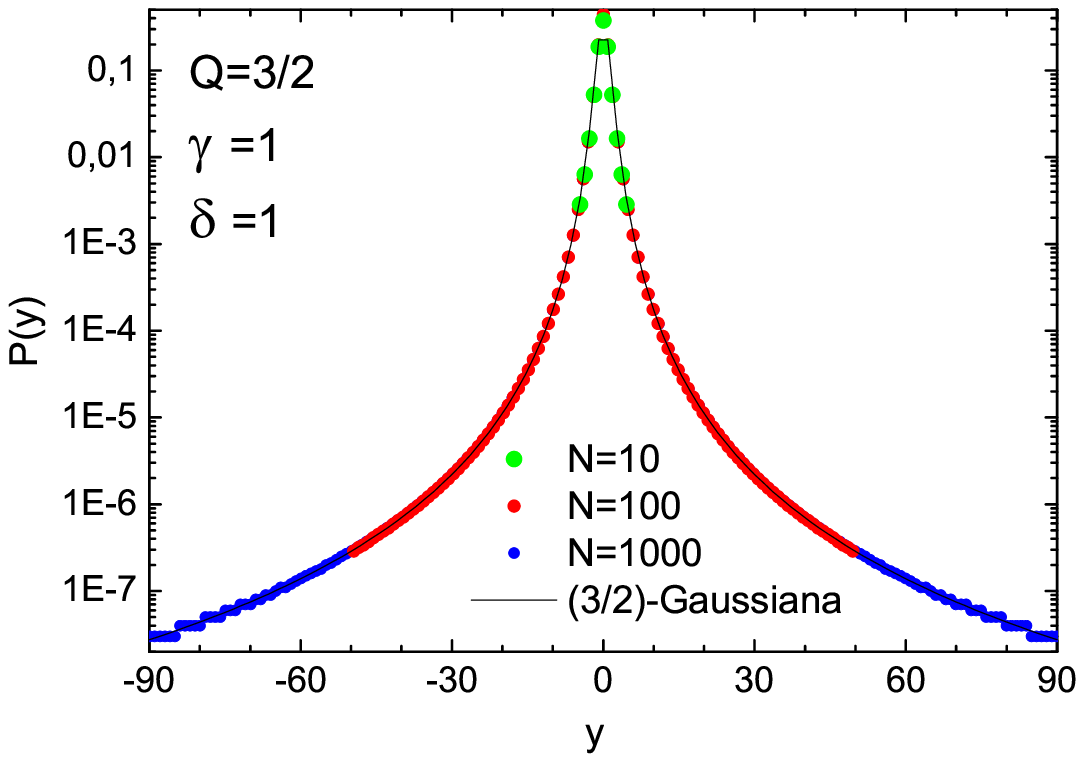}
  \includegraphics[width=6.3cm,angle=0]{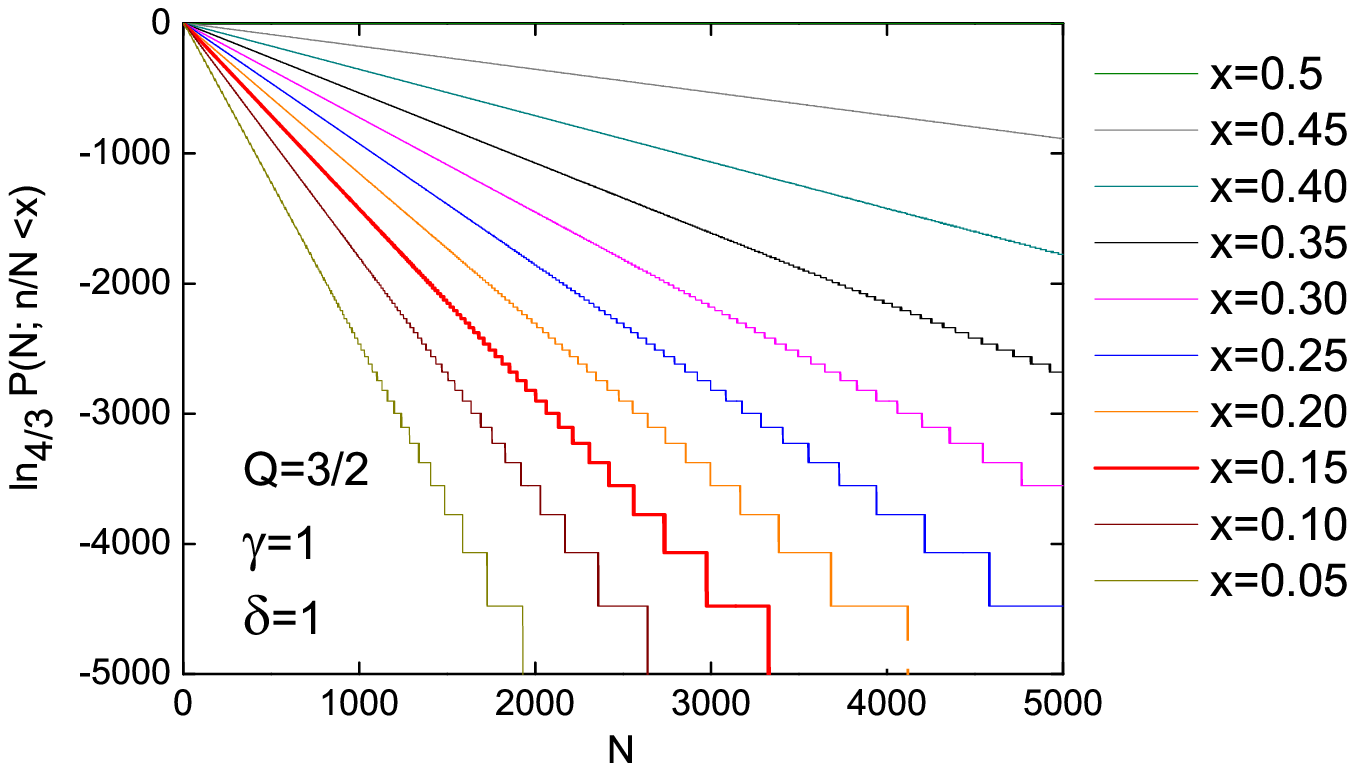}
}\vspace{-0.5cm}
\caption{\small {Illustrative histograms of the discretized model (left column) and their corresponding distributions $P(N;\,n/N<x)$, in semi $q$-log representation (right column). The cases $\gamma= 0$ and $\gamma=1$ do not yield $q$-Gaussians with $q>1$ because  the discretization never achieves the desired continuous limit (indeed, $\Delta_N=\delta$ and $\Delta_N/N \sim \delta$ respectively). The case $Q=1$ yields a Gaussian with a specific discretization, which only in the $N \to\infty$ coincides with that of the independent-coin model.\label{histogramModel}}
 }
\end{figure*}

The model is fully determined by $(Q, \,\gamma, \delta)$.
Fig.~\ref{histogramModel} (left column) shows how the distributions
$P(y)=\frac{N}{\Delta_N} p_{N,n}$ approach the corresponding $Q$-Gaussians while $N$ increases.

\begin{figure}
\begin{center}
\includegraphics[width=8cm]{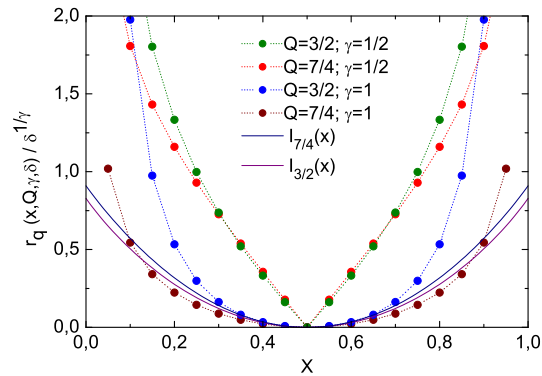}
\end{center}\vspace{-1cm}
\caption{Rate function $r_q(x)$ corresponding to typical values of $(Q, \gamma)$. The continuous curves correspond to
$I_Q(x)$ (Eq. (\ref{Iqx}) with $q \to Q$).\label{fig_rate}}\vspace{-0.5cm}
\end{figure}

We observe that, for each pair of values ($Q$, $\gamma$), $P(N; \, n/N <x)$ presents  a $q$-logarithmic decay (see Fig~\ref{histogramModel}, right column), i.e.,
\begin{equation}
P(N; \,n/N<x) \simeq e_q^{- N\,r_q(x)} \;\;\;(0 \le x \le 1/2)\,,
\label{PNq}
\end{equation}
where
\begin{equation}q = \frac{Q -1}{\gamma (3 - Q) } + 1 \;\;\;(0<x \le1/2; \,\forall \delta) \,,
\label{q1}
\end{equation}
hence,
\begin{equation}\frac{1 }{\gamma (q - 1)} = \frac{2}{Q - 1} -1 \;\;\;(0<x \le1/2; \,\forall \delta) \,.
\label{q2}
\end{equation}
We see that, for $Q=1$ hence $q=1$, Eq. (\ref{PNq}) recovers Eq. (\ref{PN1}). For $Q>1$, we have $q>1$, consequently $P(N; \,n/N<x) \propto 1/N^{1/(q-1)}$, i.e., {\it a power law instead of exponential}. We also verify that, the value of $q$ for $x=0$, noted $q_{[x=0]}$, differs (possibly due to a boundary effect) from the value corresponding to $0<x \le 1/2$, noted $q_{[0<x \le 1/2]}$ and given by Eqs. (\ref{q1}) or (\ref{q2}). For all $(Q,\gamma, \delta)$ we have that
\begin{equation}
q_{[x=0]}= 2 - \frac{1}{q_{[0<x \le 1/2]}} \,.
\end{equation}
The rate function satisfies, for $0 < x \le 1/2$,
\begin{equation}
r_q(x;\, Q;\, \gamma; \, \delta)=r_q(x;\, Q;\, \gamma; \,1)\, \delta^{1/\gamma} \;\;\; (\delta>0) \,,
\end{equation}
where $r_q(x;\, Q;\, \gamma; \,1)$ depends on the model parameters $(Q, \gamma)$, as illustrated in Fig. \ref{fig_rate} [notice that Eq. (\ref{Delta}) yields $\Delta_N \sim (N\delta^{1/\gamma})^\gamma$ for $N>>1$]. In all cases, due the above mentioned boundary effect, $r_q(0) < \lim_{x\to 0} r_q(x)$. For comparison purposes we have also represented, in this same figure, $I_Q(x)$ as given by Eq. (\ref{Iqx}). We verify that, although it is of the same order of magnitude as $r_q(x)$, it does not coincide with the numerical results from Fig. \ref{histogramModel} (neither the corresponding $I_q(x)$'s, not shown in the figure, coincide). This cannot be considered as surprising since the present model includes, for $Q>1$, nontrivial correlations between the $N$ random variables, which have not been taken into account in the calculation of (\ref{Iqx}). It is however remarkable that the exponent of the $q$-exponential (\ref{PNq}) remains extensive (i.e., proportional to $N$) for all values of $Q$. Since the nature of this exponent is entropic, this results naturally reinforces the approach currently adopted in nonextensive thermostatistics, where, in the presence of strong correlations, one expects a value of the index $q$ to exist such that $S_q$ preserves the extensivity it has in the BG theory \cite{GellMannTsallis2004,CarusoTsallis2008,Tsallis2009,SaguiaSarandy2010}.

The present study opens the door to a $q$-generalization of virtually many, if not all, of the classical results of the theory of large deviations. In this sense, the present effort points a path which would be parallel to the $q$-generalization of the classical and L\'evy-Gnedenko Central Limit Theorems \cite{UmarovTsallisSteinberg2008}. Indeed, the present results do suggest the mathematical basis for the ubiquity of $q$-exponential energy distributions in nature, just as the $q$-generalized Central Limit Theorem suggests the ubiquity of $q$-Gaussians in nature.
\vspace{-0.3cm}

\section*{Acknowledgments}\vspace{-0.3cm}
We acknowledge useful conversations with R. Bissacot, T.C.P. Lobao, S.T.R. de Pinho, H. Touchette and M.E. Vares, as well as financial support by DGU-MEC (Spanish Ministry of Education) through Project PHB2007-0095-PC.

\end{document}